# Behavior of the Dielectric and Pyroelectric Responses of Ferroelectric Fine-Grained Ceramics

Oleksandr S. Pylypchuk[1], Serhii E. Ivanchenko[2], Mykola Y. Yelisieiev[1,3], Andrii S. Nikolenko[1,3], Victor I. Styopkin[1], Bohdan Pokhylko[2], Vladyslav Kushnir[2], Denis O. Stetsenko[1], Oleksii Bereznykov[1], Oksana V. Leschenko[2], Eugene A. Eliseev[2], Vladimir N. Poroshin[1], Nicholas V. Morozovsky[1*], Victor V. Vainberg[1†], and Anna N. Morozovska[1‡]

[1] Institute of Physics, National Academy of Sciences of Ukraine,

46, pr. Nauky, 03028 Kyiv, Ukraine

[2] Frantsevich Institute for Problems in Materials Science, National Academy of Sciences of Ukraine

Omeliana Pritsaka str., 3, Kyiv, 03142, Ukraine

[3] Lashkarev Institute of Semiconductor Physics, National Academy of Sciences of Ukraine,

41, pr. Nauky, 03028 Kyiv, Ukraine

## Abstract

We revealed the specific temperature behavior of the complex dielectric permittivity and unusual frequency dependences of the pyroelectric response of the fine-grained ceramics prepared by the spark plasma sintering of the ferroelectric $BaTiO_3$ nanoparticles. The real part of the relative dielectric permittivity sharply increases with the temperature rise; then it saturates at giant values ($\sim 5 \cdot 10^5$) and remains at the plateau in the broad temperature range. The dielectric losses have a pronounced maxima in the region of the steep increase of the dielectric permittivity. The temperature dependences of the electro-resistivity indicate the frequency-dependent transition in the electro-transport mechanisms between the lower and higher conductivity states accompanied by the maximum in the temperature dependence of the loss angle tangent. The pyroelectric thermal-wave probing revealed the existence of the spatially inhomogeneous counter-polarized ferroelectric state at the opposite surfaces of the ceramic sample. We described the temperature behavior of the giant dielectric response and losses using the core-shell model for ceramic grains, effective medium approach and Maxwell-Wagner approach. The superparaelectric-like state with a giant dielectric response may appear due to the internal barrier-layer capacitance effect, while the step-like thermal activation of localized polarons in the semiconducting grains is not excluded. The elucidation of the state microscopic origin requires measurements in the frequency range above 1 MHz.

---

[*] corresponding author, e-mail: nicholas.v.morozovsky@gmail.com
[†] corresponding author, e-mail: viktor.vainberg@gmail.com
[‡] corresponding author, e-mail: anna.n.morozovska@gmail.com

# I. INTRODUCTION

The barium titanate (BaTiO$_3$) is a classical ferroelectric material [1], whose ceramics are widely used for a variety of applications, such as positive temperature coefficient thermistors, (ferroelectric posistors), pyroelectric detectors [2] and multilayer ceramic capacitors (**MLCC**) [3]. The problem of BaTiO$_3$ (**BTO**) application to ceramic capacitors requires high relative dielectric permittivity values, as high as 10$^5$, and is closely connected to the problem of the fine-grained BTO ceramics sintering and production. Traditional hot-pressing sintering (**HPS**) technologies, which use relatively high sintering temperatures (from 1200 to 1450 °C) and times (2 to 5 hours), result in BTO ceramics with a relative density of about 80% and developed granular structure. However, resulting dielectric parameters of HPS BTO ceramics are not enough for MLCC applications.

In accordance with modern requirements, it is desirable to minimize the energy consumption and time of the ceramic sintering procedure. These requirements can be fulfilled using the spark plasma sintering (**SPS**) method [4, 5, 6], which allows to obtain colossal and frequency stable permittivity of BTO nanoceramics [7], has been adapted for the product manufacturing [8] and mass production [6]. SPS allows to achieve high density and suppression of ceramic grain growth [9, 10]. In particular, the mechanical activation synthesis and SPS at 1112°C for 3 minutes under 50 MPa leads to the BTO nanoceramics with relative density of 98%, which dielectric constant is as high as $3.5 \cdot 10^5$ and loss angle tangent is about 0.07 at the frequency of 1 kHz [7].

The SPS is superior to the HPS in terms of the temperature control, the control quality of grain boundaries and fine-crystalline structure. However, the SPS uses an electrical current between particles placed in a graphite die. The presence of the conducting graphite inclusions, which concentration changes from the ceramic pellet surface towards its depth, can influence strongly on the electrophysical properties of the ceramics. Often the surface layer with the highest concentration of graphite inclusions is eliminated as contamination.

Several seminal papers [11, 12, 13, 14] report about the giant (or even colossal) dielectric permittivity of the BTO nanoceramics prepared by the SPS. Many other non-ferroelectric inhomogeneous and disordered systems also reveal the giant dielectric permittivity behavior at low frequencies (see e.g., [15, 16, 17]). It was shown [15-17] that the permittivity behavior is caused by the mesoscopic inhomogeneities of the electrical conductivity, e.g., between the grains and grain boundaries, known as the internal barrier-layer capacitance (**IBLC**) effect, as well as by the inhomogeneous layers between the electrodes and the sample, known as the surface barrier layer capacitance (**SBLC**) effect. To observe the IBLC effect responsible for the giant permittivity at low frequencies, the inter-grain conductivity should be essentially smaller than the grain conductivity, since the grain boundaries are macroscopically percolated in a dense ceramic and the grain bulk volumes are separated by the grain boundaries [11-17]. Percolation of the low-conductivity components through the sample is consistent with the giant permittivity effects in the core-shell systems, as it follows from the effective medium approximation (**EMA**) models discussed in the papers [15-17], as well as in the Petzelt et al. [18] and Rychetský et. al. [19] works.

Han et al. [13] concluded that the origin of colossal permittivity in BTO nanoceramics is due to the synergy of IBLC effect, interfacial polarization at the interior of insulating grain boundaries, and polaron hopping in semiconducting grains with a large number of induced charge carriers. They estimated that the contribution of the interfacial permittivity ($\sim 3 \cdot 10^4$) appeared about 12.13% and the contribution of polaron permittivity ($\sim 1.7 \cdot 10^5$) appeared about 73%. In 2019 Liu et al. [20] observed the colossal dielectric permittivity in the ferroelectric relaxor-like $CaCu_3Ti_4O_{12}$ ceramics. Using the Maxwell-Wagner approach [21] for inhomogeneous semiconducting media, they reveal that the giant dielectric response can appear due to the thermal activation of localized polarons [20]. The polaron hopping appear in semiconducting grains, where the charge carriers are able to move inside (but not beyond the grain boundaries), resulting in the Maxwell-Wagner effect, which significantly enhances the dielectric permittivity [20].

In this work we observed the giant dielectric permittivity and unusual frequency dependences of the pyroelectric response of the ferroelectric fine-grained ceramics prepared by SPS of the BTO nanoparticles with the average size of 25 nm. We assume that the superparaelectric-like giant dielectric response may appear due to the IBLC effect [15-17], as well as the thermal activation of localized polarons [13, 20] may be considered. Notably that the superparaelectric relaxor state has been observed in the ferroelectric $BiFeO_3$-$BaTiO_3$-$SrTiO_3$ films [22].

The paper is structured as follows: **Section II** contains the experimental results and their theoretical analysis; and **Section III** contains a summary. The details of samples preparation and characterization are listed in **Supplemental Materials** [23].

## II. EXPERIMENTAL RESULTS AND THEIR THEORETICAL ANALYSIS

We studied the polar, dielectric and pyroelectric properties in the BTO ceramics prepared by the HPS at 1250 °C at sintering time 5 hours, and the BTO ceramics prepared by the SPS at 1100 °C for 5 minutes, the heating rate is 400°C/minute (see **Appendix A** [23] for details). For both ways of sintering, we used the same BTO nanoparticles with the average size of 25 nm. The silver (or indium) contacts were made using the method of thermal evaporation in the vacuum deposition aggregate VUP-5M.

### A. Raman and SEM Studies

The Raman spectra of the BTO ceramics prepared by the HPS are shown in **Fig. 1**. The HPS BTO ceramics exhibits five Raman-active phonon modes located at 185, 258, 307, 519 and 718 cm$^{-1}$ and four dips located at 180, 297, 457 and 680 cm$^{-1}$. The mode positions are close to those known for calcined BTO nanopowders (185, 260, 306, 515 and 715 cm$^{-1}$) [7] and BTO-Ag nanocomposites (180, 260, 300, 514 and 712 cm$^{-1}$) [24]. Similar to Ref.[7], we regard that the Raman peaks at 180, 258 and 519 cm$^{-1}$ correspond to the cubic phase of BTO, and the peaks at 307 and 718 cm$^{-1}$ correspond to the tetragonal phase of BTO. Pronounced shape asymmetry of the 519 cm$^{-1}$ peak can be associated with overlapping of true $A_1$(TO) and E(TO) modes.

The weak intensity of the Raman spectra labeled as "Pwd p1" and "Pwd p2", corresponding to the different point of the powder sample, in comparison with "Tab p1" and "Tab p2", corresponding to the different point of the HPS ceramic sample, is due to the fact that these spectra correspond to the initial nanocrystalline $BaTiO_3$ powders, which differ significantly from the sintered ceramics in terms of grain size and crystalline quality, as well as have significantly lower density (that is critical for the Raman signal formation). It should be noted that the Raman spectra of the initial powders are characterized not only by the lower intensity of the phonon bands but also by their greater full width at half maximum (FWHM) and distinct shape. According to the literature data [25, 26, 27], such behavior of the Raman spectra may be attributed to the size effects due to changes in the lattice constant and the symmetry of the crystalline structure with decreasing in crystallite and grain size, as well as to disordering effects.

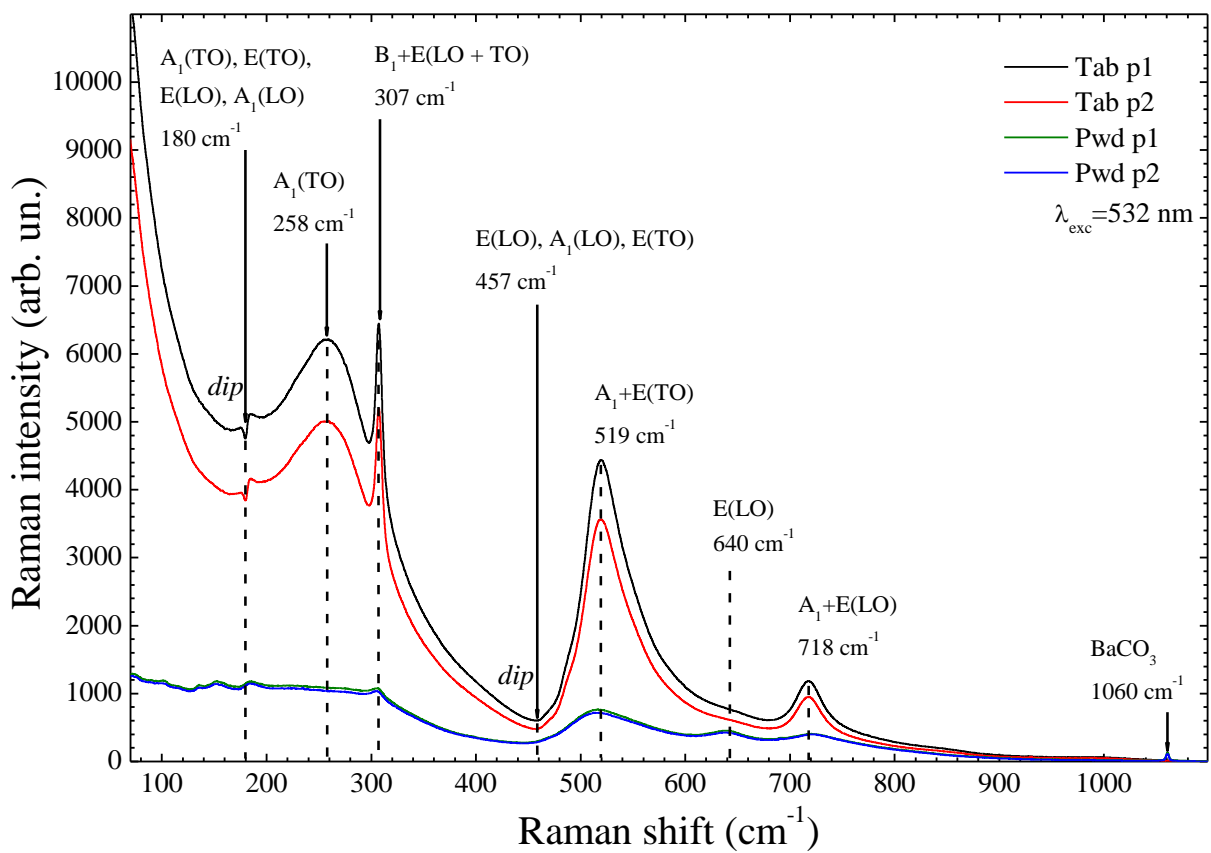

**FIGURE 1**. Raman spectra of the BTO ceramics prepared by the HPS. The abbreviation "Tab" is for the spectra of tablet of the HPS ceramic sample, the abbreviation "Pwd" is for the spectra of nanopowder sample,

which was used for the HP sintering of ceramics. The abbreviations "p1" and "p2" correspond to the characteristic spectra measured in different points of the samples. Symbols near the peaks and dips correspond to the classification of the Raman-active modes ($A_1$, $B_1$ or E, or their sum), symbols "TO" and "LO" (in round brackets) mean the transverse and longitudinal optic modes, respectively.

The scanning electron microscopy (SEM) images of the SPS ceramics revealed that the number of graphite inclusions (shown by black spots in **Fig. 2(a)**) decreases in two times from the surface towards the middle of the pellet (see **Appendix A** [23] for details). According to SEM observations we prepared several samples from the SPS ceramic pellet for further studies, shown in **Fig. 2(b)**, where different samples have different concentrations of the graphite inclusions. The whole pellet has dimensions 6.2x5.6x2.4 mm$^3$, the dimensions of the sample # 1 are 5.6x2.6x2.0 mm$^3$, and the three thin slices with high and low concentrations of graphite inclusions, designated as the samples # 2, 3, and 4, have dimensions 4.5x2.0x0.2 mm$^3$.

The Raman spectra of the BTO ceramics prepared by the SPS and recorded in different points p1-p9 of the prism-shaped sample are shown in **Fig. 2(c)** and **2(d)**. The Raman spectra of the SPS BTO ceramics are very similar to those of HPS ceramics: one can see five Raman-active phonon modes located at 183, 258, 307, 519 and 718 cm$^{-1}$ and four dips located at 180, 297, 457 and 680 cm$^{-1}$ (compare with **Fig. 1**). However, the Raman spectra of the SPS BTO ceramics contains the carbon bands D (1330 cm$^{-1}$) and G (1610 cm$^{-1}$), which are clearly seen for p5, p5' and p8 points in **Fig. 2(d)** corresponding to the dark spots.

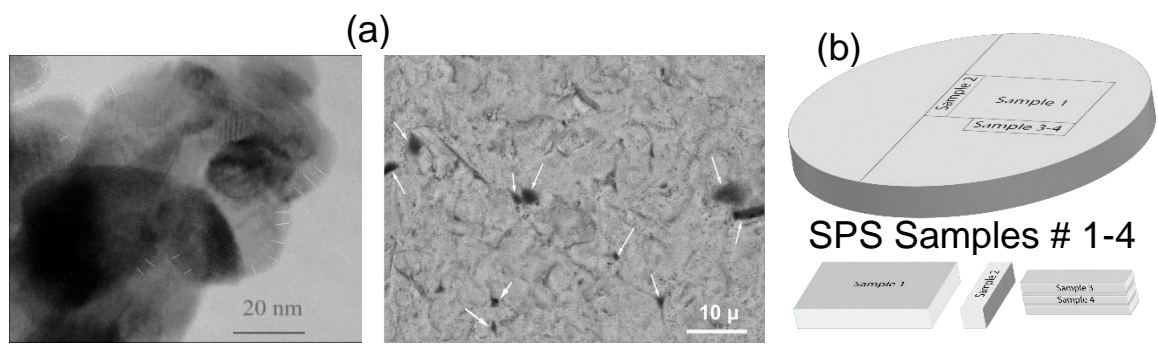

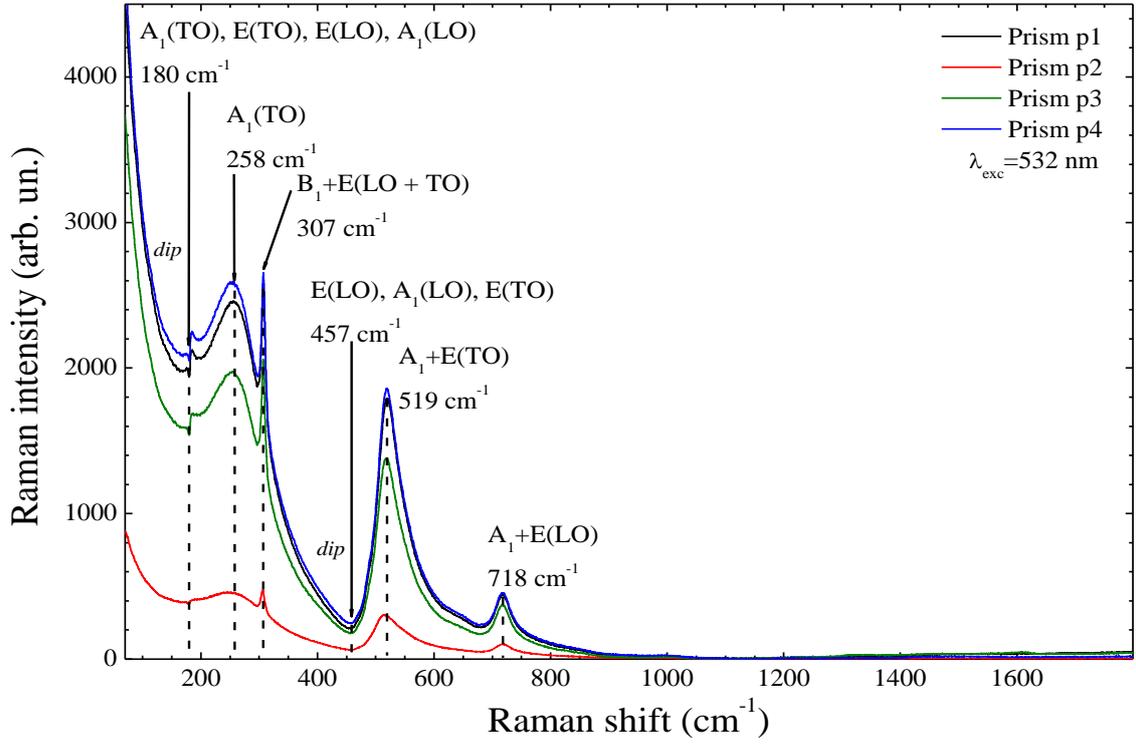

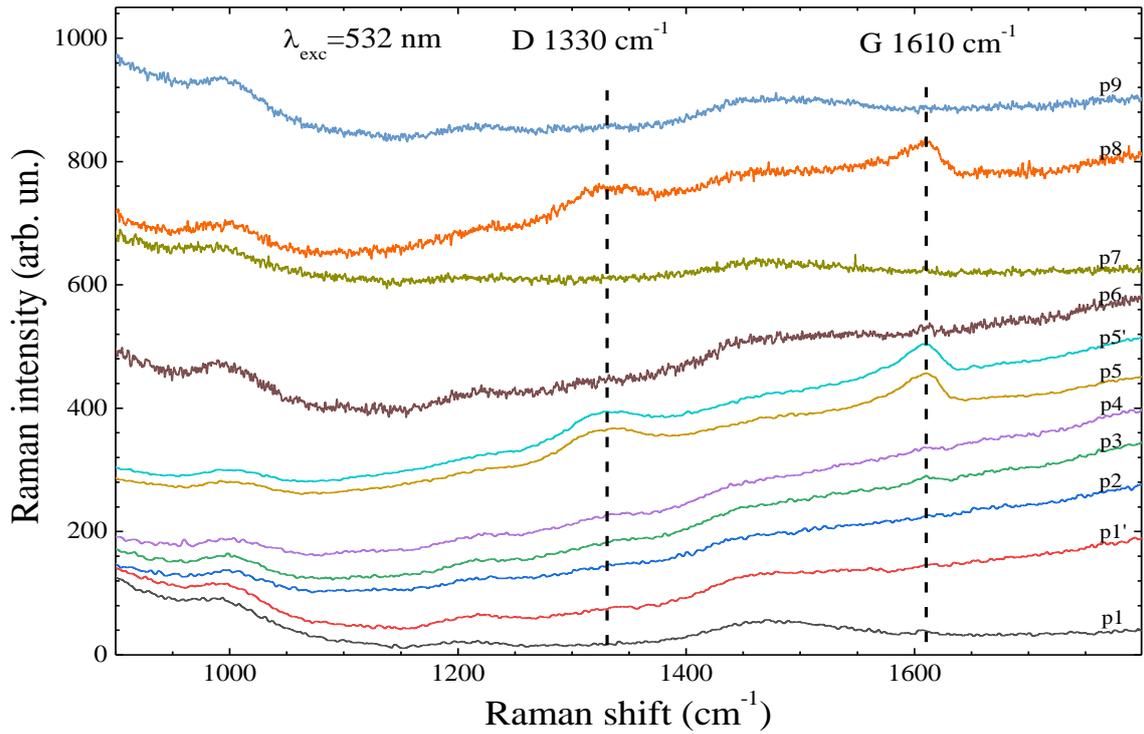

**FIGURE 2**. **(a)** TEM image of the grains and Z-contrast SEM image of the SPS pellet edge. White arrows point on the black graphite inclusions. **(b)** Cutting diagram of the SPS BTO pellet. **(c)** Raman spectra of the BTO ceramics prepared by the SPS. **(d)** The part of the Raman spectra **(c)** recorded in different points p1-p9 of the prism-shaped sample of the SPS BTO ceramics.

## B. Dielectric Permittivity and Losses

The temperature dependences of the dielectric permittivity ($\varepsilon$) and loss tangent ($tg\delta$) of the HPS BTO ceramic pellet are shown in **Fig. 3**. It is seen that the permittivity, measured at 1, 10 and 100 kHz, has a sharp peak at approximately 125°C, which corresponds to the paraelectric-ferroelectric phase transition temperature in a bulk BTO single-crystal, and a relatively smooth maxima at about 20°C, which corresponds to the orthorhombic-tetragonal phase transition temperature and is also close to the corresponding transition temperature in a bulk BTO single-crystal. The position and value of the dielectric permittivity maximum are almost frequency-independent. The losses have a relatively sharp minima at 125°C, and a pronounced maximum at 20°C. The temperature behavior is typical for the dense HPS BTO ceramics with ferroelectric grains.

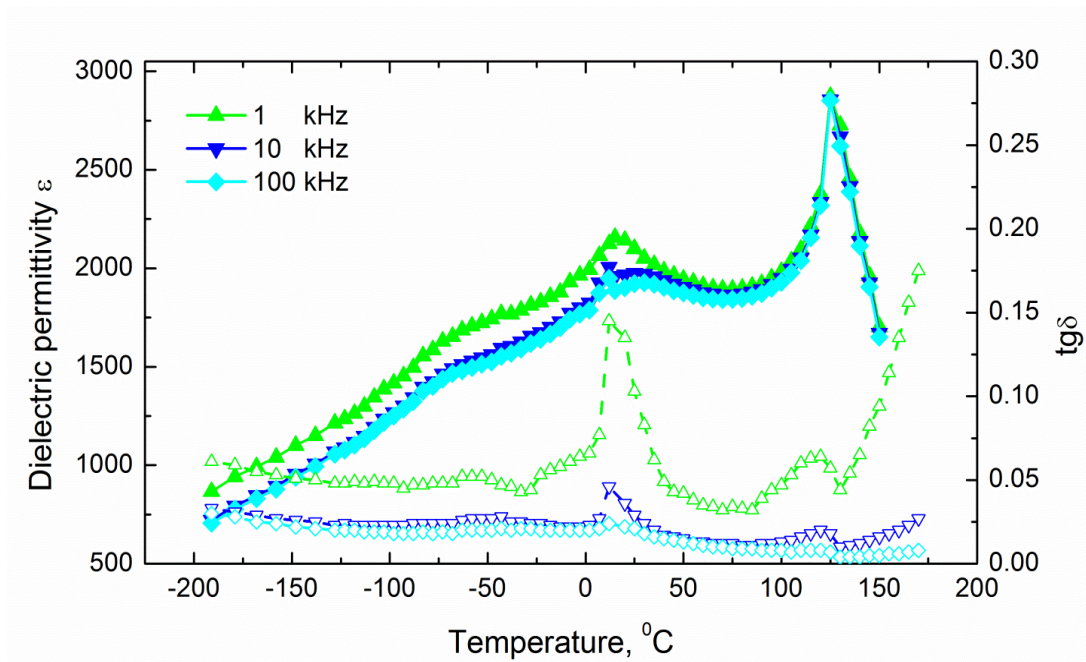

**FIGURE 3**. The temperature dependences of the dielectric permittivity (ε, filled symbols) and the dielectric losses (tgδ, empty symbols) of the BTO ceramics prepared by the HPS.

The temperature dependences of the dielectric permittivity for the SPS samples # 1 - 4 are shown in **Fig. 4(a)** - **4(d)**, respectively. The permittivity, measured at frequencies from 100 Hz to 100 kHz, has no pronounced maxima in the studied temperature range. Instead, the dielectric permittivity increases very strongly (by 4 orders of magnitude) under the temperature increase from -200°C to +100°C. The increase rate is the highest between -150°C and -50°C when the permittivity reaches the values ~ (2 - 8)·$10^5$, then the increase rate becomes much slower due to the permittivity saturation. The pronounced permittivity saturation appears above -75°C (for 100 Hz, 120 Hz and 1 kHz), above -50°C (for 10 kHz) and above +25°C (for 100 kHz). The slow increase of the dielectric permittivity (instead of the saturation) is observed above +75°C at 1 kHz; other curves continue to saturate in the studied temperature range. The temperature and frequency dependences of the dielectric permittivity for the samples # 1 - 4 are qualitatively similar. The main quantitative differences are in the increase rate and saturation values of the dielectric permittivity, which can vary by one order of magnitude. Unfortunately, these measurements were limited in the temperature range below the bulk Tc (namely less than (100 – 110)$^0$C), because of the sharp, for some samples uncontrolled, increase in the active part of the conductivity and the output of the measuring device (LRC-meter) beyond the permissible technical range. For some samples a breakdown-like effect occurred.

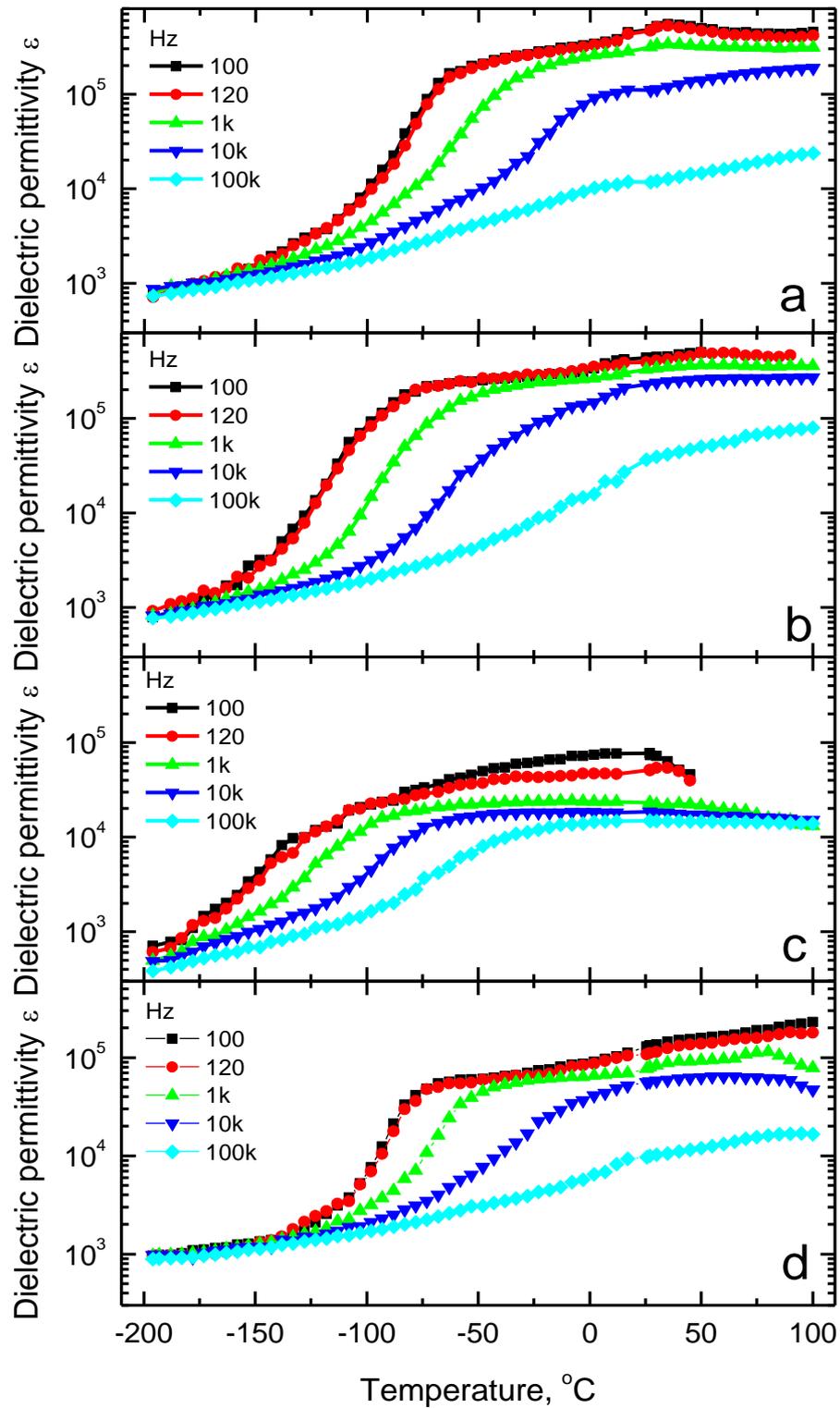

**FIGURE 4**. The temperature dependences of the dielectric permittivity measured for the samples # 1 **(a)**, #2 **(b)**, #3**(c)** and #4 **(d)** at frequencies from 100 Hz to 100 kHz.

The temperature dependences of the $tg\delta$ measured for the SPS samples # 1 - 4 are shown in **Fig. 5(a)** - **5(d)**, respectively. The temperature dependence of the $tg\delta$ is non-monotonic for all considered frequences, with relatively sharp maxima at -150°C (for 100 Hz, 120 Hz and 1 kHz), -100°C (for 10 kHz) and a wide peak at -25°C (for 100 kHz). The temperature positions of the maxima are located in the middle of the region, where the dielectric permittivity increases rapidly. For the temperatures above the maximum, the losses start to increase rapidly with the temperature increase (for 100 and 120 Hz) or more slowly (for 1 kHz). We associate the increase with the likely thermal breakdown. The increase (and thus the breakdown) is absent (or occur at much high temperatures outside the studied range) for 10 kHz and higher frequencies.

For most samples (#1, 2 and 4) the maximum of $tg\delta$ shifts to the higher temperatures and its height very slightly increases with the increase of frequency $\omega$ from 100 Hz to 100 kHz. We observed $tg\delta \sim \omega^\alpha$, where $\alpha$ is a small positive number. Since our measurements correspond to the parallel connection in the equivalent circuit, $tg\delta \sim 1/\omega CR$, we may suggest the measured frequency dependence of $tg\delta$ is determined by a mixed connection of the equivalent dielectric capacitors and resistors.

The maximum in the temperature dependence of the $tg\delta$ is a reliable signature of the change in the mechanism of the sample electrical transport and dielectric response. Active losses, caused by the energy release of the active part of the resistance, do not increase in the maximum region, but the reactive component, which is a combination of the volume capacitances (~permittivities), changes sharply. In result the ratio of the active and reactive components also changes sharply leading to the maximum in the $tg\delta$. Thus, the maximum of the losses accompanying the sharp changes in the dielectric permittivity is a fingerprint of the electrical transport mechanism change entire the sample volume.

The temperature and frequency dependences of the losses for the samples # 1, 2 and 4 are qualitatively similar. The main quantitative differences are in the positions of the $tg\delta$ maximum, and the value of losses in their maximum. The breakdown can occur at frequencies below 10 kHz. Let us underline, that the sample # 3, where $tg\delta$ has no maxima, reveals the lowest resistivity and the lowest dielectric permittivity (less than $10^5$) in the high temperature range. Also, the sample has the highest losses among the studied samples, which become very high starting from the lowest temperature. This behavior can be related with the sharp changes of polarizability at high concentration of the graphite contamination in the surface part of the pellet.

Note that, though technical reasons do not allow us to perform reliable dielectric measurements above 100°C, we show in **Figs. 4 - 5** the dielectric response measured in the temperature range from -200°C to + 100°C, which contains, on our opinion, all significant changes of the response.

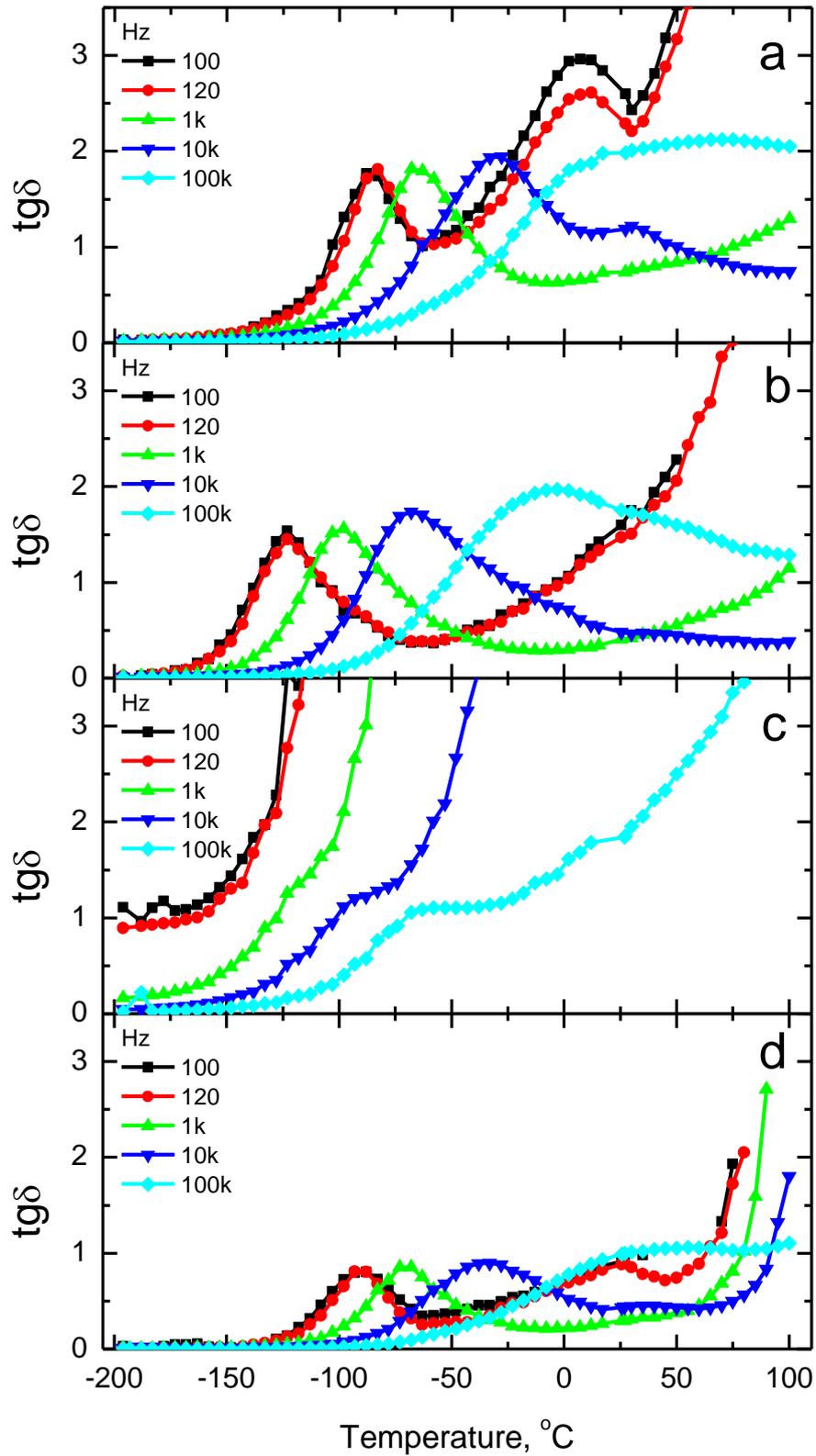

**FIGURE 5**. The temperature dependences of the loss angle tangent measured for the samples # 1 (**a**), #2 (**b**), #3(**c**) and #4 (**d**) at frequencies from 100 Hz to 100kHz.

## C. Electric Transport Characteristics

Shown in **Fig. 6** are the temperature dependences of the SPS samples # 1-4 resistivity measured in the direct current (DC) regime in the range from the room temperature down to the liquid nitrogen temperature. It is seen that the dependences have a semiconductor activation type behavior in the Arrhenius coordinates with the activation energies in the range from 120 to 290 meV. The thermo-emf vs $\Delta T$ dependence for the sample #2 measured at 298 K is shown in the inset. The thermo-emf sign corresponds to the electron type conduction. The dependence has two linear parts with the differential thermo-emf of 436 and 606 µV/grad, respectively.

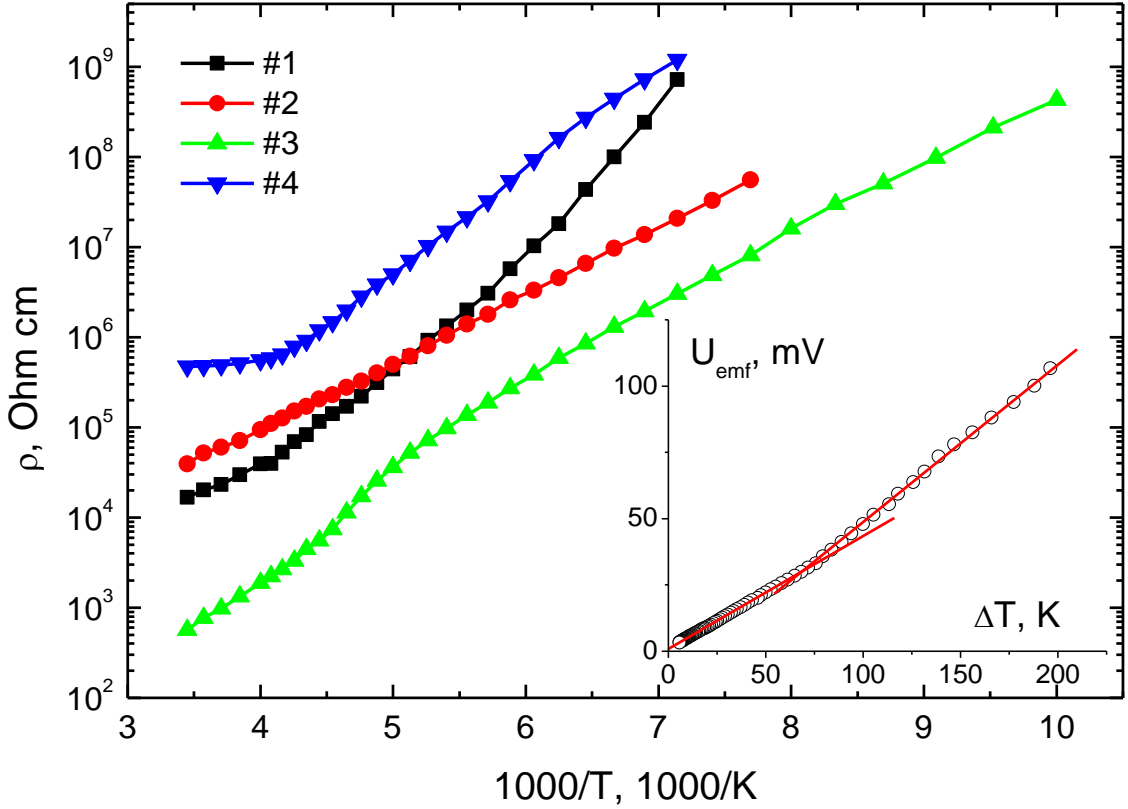

**FIGURE 6.** The temperature dependences of the resistivity measured for the samples # 1**,** 2**,** 3 and 4 under the small electric field. Inset: the dependence of thermo-emf vs $\Delta T$ for the sample #2, T= 298 K.

The temperature dependence of conductivity with high activation energies are characteristic for semiconductors and may be explained by the models in which potential barriers for movement of charge carriers are formed. The small polaron hopping conduction models [28, 29, 30] are inherent to macroscopically homogeneous samples. For considered inhomogeneous samples more appropriate maybe the models based on the concept of the inhomogeneous medium, also leading to the hopping conduction [31]. Also, the electron-type conductivity in the considered case may include the conductivity due to the charges in the inter-grain space (while the inter-grain space is still much less conducting than the grains). The potential barriers may be caused by non-homogeneously spatially distributed charges between the grain bulks and their boundaries (strongly different conductivities), which allows us to measure only some effective characteristics. Also, the activation type behavior of the conductivity may be related to deep impurity states of the oxygen vacancies (see e.g. Ref.[32]), and the possibility requires further studies.

Hopping or other conduction types with a strong temperature dependence of resistivity applies a specific character on conduction in strong electric fields. Also, measurements of the DC current vs time have shown that all samples (more or less) are subjected to the Joule heating under the increase of the DC voltage. It is illustrated for the sample #3 in **Fig. 7(a)**, where the temporal dependences of the relative current through the sample are shown at different applied voltages. To avoid the Joule heating, the current-voltage characteristics (I-V curves) were measured in the pulsed regime with the voltage pulse duration no more than 1 ms. The I-V curves at four temperatures for the sample #3 are shown in **Fig. 7(b)**. The I-V curves are slightly super-linear and do not intercept even at the electric field strength up 2 kV/cm. This indicates that the applied electric fields are still too small to affect the energy barriers determining the conduction mechanisms. The samples #1, 2 and 4 demonstrate a similar behavior.

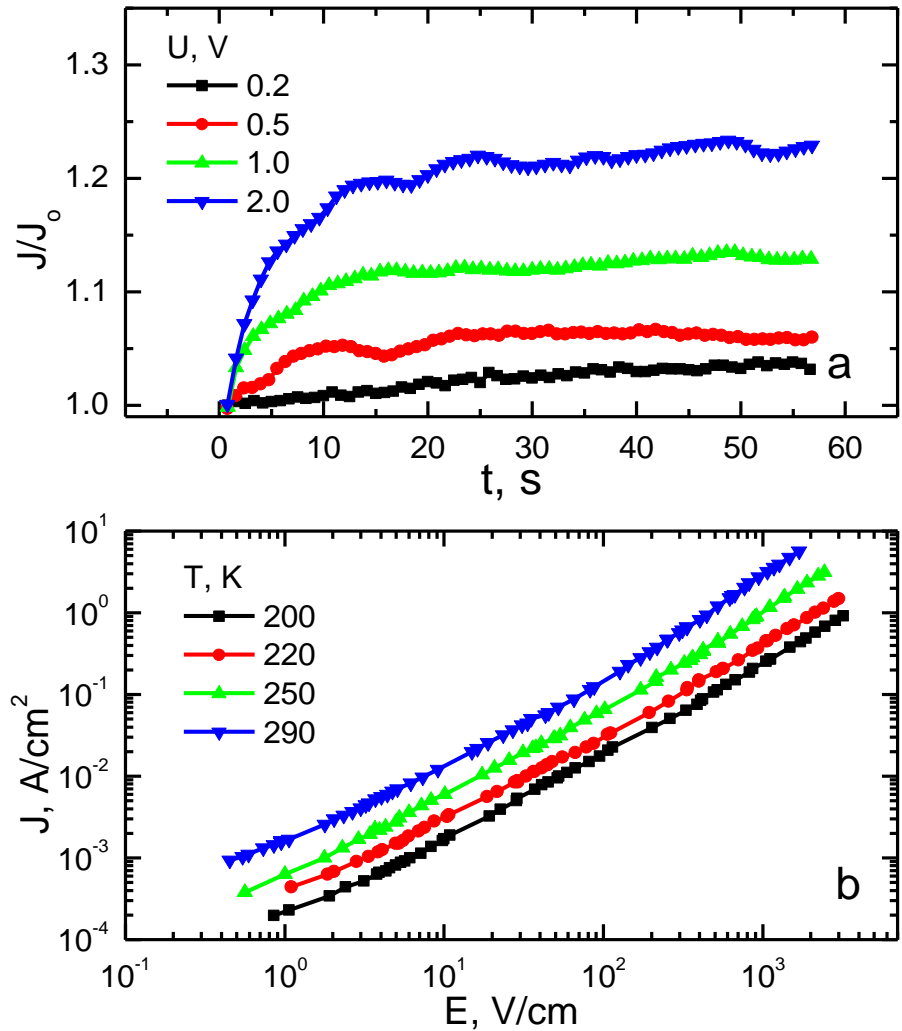

**FIGURE 7**. **(a)** Temporal dependences of the relative current through the sample at different applied voltages. **(b)** Current-voltage characteristics of the sample #3 in the pulsed regime in the temperature range (200 – 290) K.

The temperature dependences of resistivity of the samples in the alternate (AC) current regime demonstrate certain well-pronounced features compared to the DC regime. Shown in **Fig. 8** are resistivity vs temperature dependences for all four samples at frequences 100 Hz, 1, 10 and 100 kHz and the DC curves for comparison. The dependences measured in the AC regime still reveal the strong activation type resistivity vs. temperature dependences at the higher temperatures and are close to the DC curves at the lowest frequency. However, unlike the DC regime, these dependencies obviously tend to saturate at the lower temperatures at all frequencies.

The second very interesting and important feature is revealed for the samples #1, 2 and 4 with higher resistivity. The resistivity of the samples demonstrates two pronounced activation regimes in the higher temperature interval and a wave-like transition (with a region of a negative temperature slope) between them. This feature becomes less pronounced and eventually disappears for the sample #3 with lower resistivity. Such feature can give grounds to suppose the existence of two phases with their own temperature dependences of the resistivity in the AC regime. It is also evident that the phases volume ratio changes with the temperature change. The transition between the phases shifts to a higher temperature with the frequency increase. The transition temperature of resistivity and its frequency shift correspond to the temperature of the $tg\delta$ maximum.

Also, we should notice that the AC temperature dependences of resistivity have certain correlations with the temperature behavior of the dielectric permittivity (compare the temperature range of the wave-like features in **Fig. 8** with the temperature of the dielectric permittivity sharp changes in **Fig. 4**). With lowering temperature, the dielectric permittivity decreases and the resistivity increases. This can evidence the decrease in the electron-type charge concentration participating in the AC conductivity. At the same time, it is also evidenced by the fact that the frequency increase causes a decrease of the dielectric permittivity and resistivity. Concerning the strong inhomogeneity of "carbon doping" of the sintered pellet, illustrated in the SEM images in **Fig. 2(a)**, one should also notice that large carbon inclusions (visible there) likely do not present all contamination. The less inclusions may spread along all inter-grain interfaces, while different by amount in different parts of the SPS ceramic pellet. We suppose that this factor influences the DC and AC electric transport characteristics in all samples and may determine differences among those cut from different parts of the pellet.

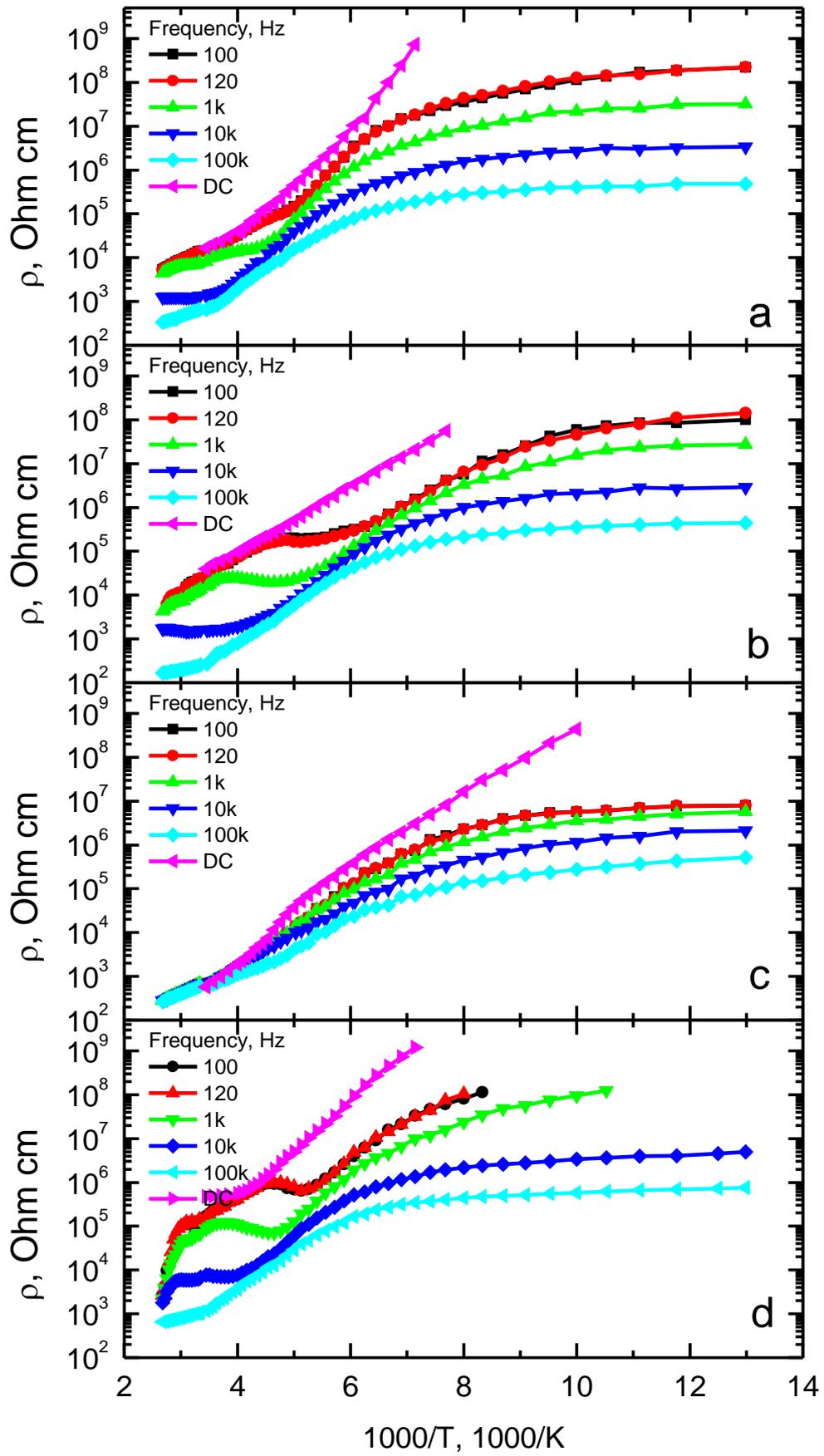

**FIGURE 8**. The temperature dependences of the samples # 1, 2, 3 and 4 resistivity measured in the AC regime (panels **a, b, c** and **d**, respectively). Violet DC curves are shown for comparison.

## D. Pyroelectric Response Measurements

To clarify the polar state of the BTO ceramics obtained by the SPS method, their pyroelectric response has been studied. For pyroelectric studies, the photo-thermo-modulation pyroelectric method [33, 34] was used. In the method, the temperature of the sample changes as $T(t) = T_0 + \Delta T \sin(2\pi f_m t)$, where $\Delta T \ll T_0$. The temperature change $\Delta T$ occurs under the influence of a sinusoidally modulated IR radiation flux with intensity $W(t) = \Delta W(1 + \delta \cdot \sin(2\pi f_m t))$, where $f_m$ is the modulation frequency and the parameter $\delta$ can vary in the range $0 < \delta \leq 1$. To create the maximal modulation amplitude, the thermal flux of the IR-diode is sinusoidally modulated by the electric voltage with the frequency $f_m$. Since the flux is absent when the diode is closed, $\delta \approx 1$ in the considered case. The measured value is the pyroelectric response $U_\pi$, caused by the temperature change of the spontaneous polarization $P_s(T)$ of the polar active material. The value of $U_\pi$ is proportional to the value of the pyroelectric coefficient $\gamma = dP_s(T)/dT$ [35].

The dynamic thermal excitation allows to analyze the amplitude-frequency, $U_\pi(f_m)$, and phase-frequency, $\varphi_\pi(f_m)$, dependences of $U_\pi$ in a wide frequency range. Under such conditions, it is possible to measure $U_\pi$ in the pyroelectric current mode, $U_\pi = U_{\pi 1}$, if $2\pi f_m R_L C_s \ll 1$ and in the pyroelectric voltage mode, $U_\pi = U_{\pi 2}$, if $2\pi f_m R_L C_s \gg 1$ (here $R_L$ is the electrical resistance of the load in the circuit of the sensitive element, $C_s$ is the electrical capacitance of the element) [36].

In the case of a uniform distribution of pyroactivity over the thickness, $U_{\pi 1,2}$ depends on the pyroelectric coefficient $\gamma$, volume heat capacity $C_\rho$ and dielectric constant $\varepsilon$ of the material and its thickness $d$ [36]. In the pyroelectric current mode, $U_{\pi 1} \propto (\gamma/C_\rho)R_L/d = \text{const}(f_m)$ and $\varphi_\pi = \varphi_{\pi 1} = \text{const}(f_m)$. Due to the resistive nature of the load in the sample circuit, $U_{\pi 1}(t)$ can be in-phase ($\varphi_{\pi 1} = 0$) or in anti-phase ($\varphi_{\pi 1} = 180°$) with the IR flux intensity $W(t)$, depending on the direction of polarization in the sample. In the pyroelectric voltage mode, $U_{\pi 2} \propto (\gamma/C_\rho \varepsilon)/f_m$, and $\varphi_\pi = \varphi_{\pi 2} = \text{const}(f_m)$. As a result of the capacitive nature of the load in the sample circuit, there is a phase shift of $\pm \pi/2$ (90° or 270°) between the $U_{\pi 2}(t)$ and the $W(t)$, depending on the polarization direction in the sample. It should be noted that the $U_{\pi 1}$ mode, chosen for the corresponding $f_m$ and $R_L$ so that $2\pi f_m R_L C_s < 1$, transforms into the $U_{\pi 2}$ mode when $2\pi f_m R_L C_s > 1$, and the transition frequency $f_{mt}$ is determined from the relation $2\pi f_{mt} R_L C_s = 1$.

Due to the relationship of the thermal diffusion length $\lambda_T = (a_T/\pi f_m)^{1/2}$ ($a_T$ is the thermal diffusivity) with $f_m$, it is possible to determine the features of the subsurface distribution of pyroelectric parameters using the thermal wave profiling [34]. Since any deviation from a uniform polarization distribution is reflected in the frequency dependences $U_{\pi 1,2}(f_m)$ and $\varphi_{\pi 1,2}(f_m)$, the corresponding $\lambda_T$-profiles of pyroelectric response $U_{\pi 1,2}(\lambda_T)$ and phase $\varphi_{\pi 1,2}(\lambda_T)$ can be obtained and analyzed [34].

The results for the SPS BTO ceramics obtained at room temperature are presented in **Fig. 9**. The upper panels show the $U_\pi(f_m)$ dependences, and the lower panels show the dependences of the phase shift $\varphi_\pi(f_m)$ between $U_\pi(t)$ and $W(t)$. The left panel of **Fig. 9** corresponds to the top surface of the BTO pellet, and the right panel of **Fig. 9** corresponds to the bottom surface.

At low frequencies, the $U_\pi$ value at one surface is significantly higher than that for the other. As $f_m$ increases the values of $U_\pi(f_m)$ of both surfaces decrease in a different way. At higher frequencies, the values of $U_\pi$ are almost the same, but corresponding values of $f_m$ are 1 kHz for one surface and 200 Hz for the other, are different (see left and right upper panels of **Fig. 9**). For both surfaces at low frequencies, the values of $\varphi_\pi$ are close to 180°, which corresponds to the pyroelectric current mode, but there is no $U_\pi(f_m) = U_{\pi 1} = const(f_m)$, characteristic of this mode (see above). At higher frequencies, $U_\pi(f_m)$ decreases with increasing $f_m$, but there is no dependence $U_\pi(f_m) = U_{\pi 2} \sim 1/f_m$, characteristic for the pyroelectric voltage mode. For both surfaces, the phase $\varphi_\pi(f_m)$ increases with increasing $f_m$ almost logarithmically (see the bottom left and right panels in **Fig. 9**). Therefore, the obtained dependences $U_\pi(f_m)$ and $\varphi_\pi(f_m)$ differ from the dependences characteristic for a homogeneous polar state.

Estimation of the thermal diffusion length $\lambda_T = (a_T/\pi f_m)^{1/2}$ for diffusivity $a_T \approx 10^{-6}$ m²/s characteristic of BTO ceramics [37], frequences $f_m$ = 10 Hz and 1 kHz gives $\lambda_T \approx 200$ μm and ~20 μm, respectively (the upper axis in **Fig. 9**). Therefore, $\lambda_T$ is big enough to carry out the thermal wave probing of the volume at low frequencies (~10 Hz), while at high frequencies (~1 kHz) $\lambda_T$ is small and the thermal wave probing of the near-surface region is successfully carried out (as shown in **Fig. 9**). Thus, the increase in $\varphi_\pi(f_m)$ with increasing $f_m$ starts from $\varphi_\pi \approx 200°$ under conditions of volume probing ($\lambda_T \sim 100$ μm) and continues under conditions of probing the near-surface layer ($\lambda_T \sim 30$ μm).

The obtained thermo-wave $\lambda_T$-profiles of pyroelectric response and phase shift of the SPS BTO ceramics sample are presented in the **Fig. 10**. The left panel of **Fig. 10** corresponds to the top surface of the ceramic pellet, and the right panel of **Fig. 10** corresponds to the bottom surface. The decrease in $U_\pi(\lambda_T)$ and increase in $\varphi_\pi(\lambda_T)$ with increase in $\lambda_T$ (**Fig. 10**) begin under the conditions of volume probing ($\lambda_T \sim 100$ μm) and continue under the conditions of probing the near-surface layer ($\lambda_T \sim 30$ μm). The maximal drop of $U_\pi(\lambda_T)$ begins after transition from the volume to the near-surface probing and correlates with the maximal rise of $\varphi_\pi(\lambda_T)$. It should noted, that such decrease of $U_\pi(\lambda_T)$ and increase in $\varphi_\pi(\lambda_T)$ may be associated with the existence not only of polar, but also thermal inhomogeneities, which are caused by graphite inclusions during the ceramic's formation by the SPS method. The specifics of the observed $\lambda_T$-profiles can be associated with a lower concentration of graphite inclusions in the volume than in the under-surface layer of the BTO ceramics.

Such $\lambda_T$-profiles of $U_\pi$ and $\varphi_\pi$ for both surfaces of the sample correspond to opposite directions of polarization under both surfaces and, therefore, the existence of an inhomogeneous counter-polarized state, like those observed in the TGS-PEO composites [34].

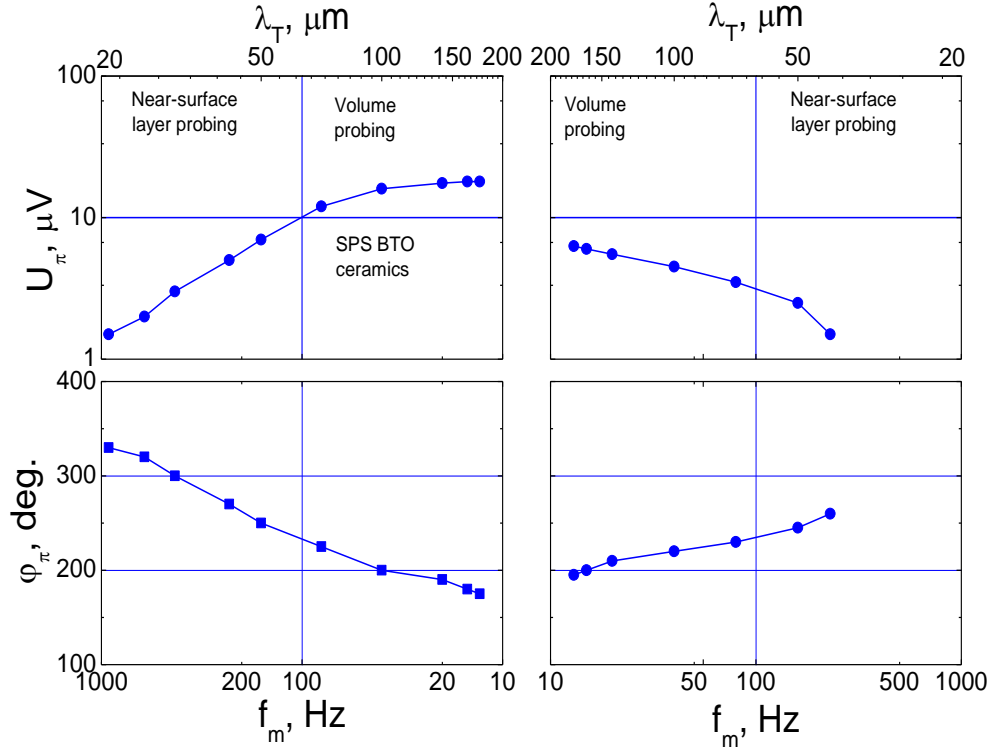

**FIGURE 9**. Dependences of the pyroelectric response amplitude (upper panels) and its phase shift (lower panels) of the SPS BTO ceramics vs. the modulation frequency of the thermal flux. The left panel of the figure corresponds to the top surface of the pellet, and the right panel of the figure corresponds to the bottom surface.

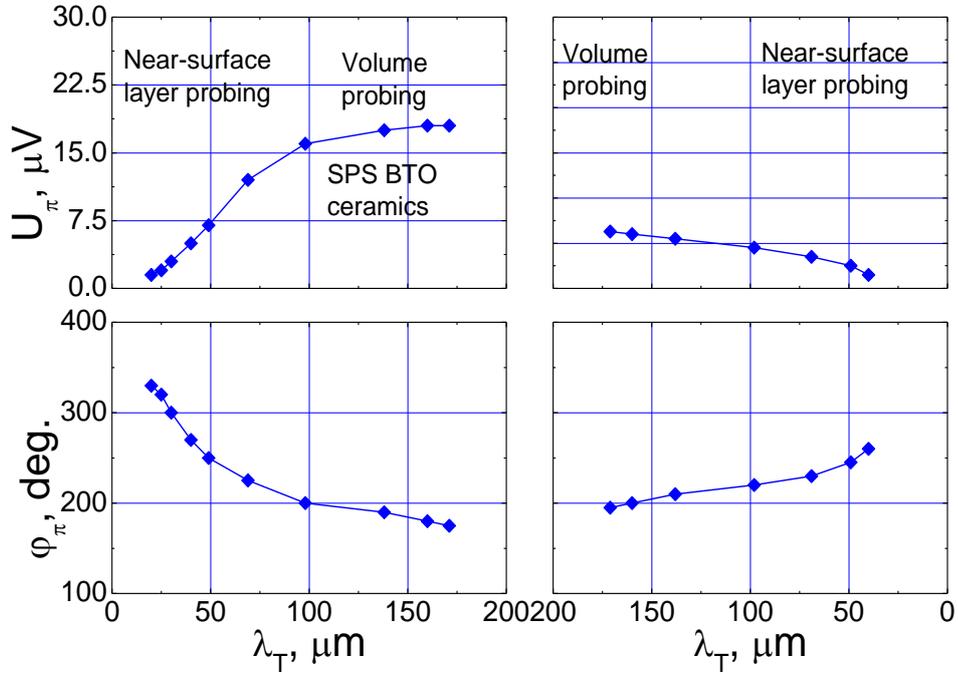

**FIGURE 10**. Thermo-wave profiles of the pyroelectric response amplitude (upper panels) and phase shift (lower panels) of the SPS BTO ceramics. The left panel of the figure corresponds to the top surface of the pellet, and the right panel of the figure corresponds to the bottom surface.

## E. Analysis of the Dielectric Response

Since the X-ray diffraction and NMR studies confirm the coexistence of the tetragonal ferroelectric and cubic paraelectric phases in the (15 – 45)-nm BTO nanopowders [38, 39] used for the preparation of the HPS and SPS ceramics, the temperature behavior of the dielectric permittivity real pat in the HPS ceramics has a pronounced Curie-Weiss behavior near the paraelectric-ferroelectric transition at about 125°C. It was shown experimentally (see e.g., Refs. [40, 41] and the review [42]) and theoretically (see e.g., Refs. [43, 44]) and the size effect can make a significant contribution to the ferroelectric transition temperature $T_{FE}$ for the grain core sizes below (15 – 30) nm.

The increase of electric current appears near the transition temperature due to the release of the trapped free carriers, as well as the pronounced maxima of the current should appear near the coercive field. For the temperatures well above the ferroelectric-paraelectric transition temperature $T_{FE}$ and/or far away from other structural transitions the current and corresponding losses should be minimal, because an "extra" charge drains through the electrodes. This well-known behavior of the carries can explain the temperature dependence of the pyroelectric current in dense HPS ferroelectric ceramics, as well as it predicts the losses maxima near the structural (i.e., rhombohedral-orthorhombic and orthorhombic-tetragonal) and paraelectric-ferroelectric phase transitions observed in the studied HPS BTO ceramics (see e.g., **Fig. 3** in this work and Fig. 6 in Ref.[14]). Note, that the classical Debye relaxation model predicts the maxima of dielectric losses and double maxima separated by the minima of the dielectric permittivity real part near $T_{FE}$ (see e.g., Eqs.(1)-(2) in Ref.[45]), which are observed in some ferroelectrics (see e.g. Fig. 1 and 2 ibidem).

In the introduction (see e.g., Refs.[13, 15, 16, 20]) we mention several mechanisms, which can lead to the giant values of the relative dielectric permittivity (>$10^5$) in the strongly inhomogeneous ferroelectric-semiconducting SPS ceramics. According to the general effective medium approach (see e.g., Refs.[16-19, 20, 46, 47, 48, 49, 50]) and the Maxwell-Wagner the ceramics can be considered as a mixture of two components representing all grain cores (the component 1), their screening shells, grain boundaries and inter-grain space with (or without) graphite inclusions (the component 2). The components are characterized by the effective dielectric permittivity and conductivity: $\varepsilon_g$ and $\sigma_g$ for grain cores, $\varepsilon_{gb}$ and $\sigma_{gb}$ for grain boundaries and inter-grains. These effective parameters can be temperature and/or frequency dependent. The complex dielectric permittivity of the components can be introduced as $\varepsilon_g^* = \varepsilon_g - i\frac{\sigma_g}{\omega}$ and $\varepsilon_{gb}^* = \varepsilon_{gb} - i\frac{\sigma_{gb}}{\omega}$. In the case when the morphology of the grains' connection is mixed, we the following expression for the effective medium permittivity is valid:

$$\varepsilon_{eff}^*(T,\omega) \approx \varepsilon_{gb}^* \left[ 1 + \frac{\mu(\varepsilon_g^* - \varepsilon_{gb}^*)}{n_g \varepsilon_g^* + (1-n_g)\varepsilon_{gb}^* - n_g \mu(\varepsilon_g^* - \varepsilon_{gb}^*)} \right], \qquad (1)$$

where the parameter $\mu$ has the meaning of the relative volume fraction of grains and $n_g$ is the depolarization factor of the grains in the direction of applied electric field (see e.g. Ref.[47]). The grains are assumed to have an ellipsoidal shape and uniformly oriented. The derivation of Eq.(1) is given in **Appendix B** [23], as well as in Refs. [46, 51] (using other designations). Similar expressions were derived by Rychetsky and Petzelt [42, 52, 53].

Equation (1) contains 6 parameters ($\mu$, $n_g$, $\varepsilon_g$, $\sigma_g$, $\varepsilon_{gb}$ and $\sigma_{gb}$) and is applicable for the mixed type of connection of insulating and/or semiconducting elements. To reproduce semi-quantitatively the observed dielectric response of the SPS ceramics one should consider $\varepsilon_g$, $\sigma_g$, $\varepsilon_{gb}$ and $\sigma_{gb} \ll \sigma_g$ as fitting frequency-dependent and temperature-dependent fitting functions. Also, we do not know the relative volume fraction of grains $\mu$ and the "effective" depolarization factor $n_g$ from the independent measurements and should treat them as fitting parameters. The fitting results with so many fitting parameters and functions do not allow us to make conclusions about the dominant mechanism of the giant dielectric response (e.g., IBLC [15, 54], or polaron activation[13, 20]). Due to the limited experimental range of the dielectric measurements (below 1 MHz), it is not possible to specify the microscopic mechanism of the grain conductivity, which could be determined from the dispersion curves measured in the higher frequency range (see e.g., Refs. [18, 55, 56]).

### III. DISCUSSION AND CONCLUSIONS

We revealed the specific temperature behavior of the complex dielectric permittivity and unusual frequency dependences of the pyroelectric response of the fine-grained ceramics prepared by the SPS of the ferroelectric BTO nanoparticles with the average size 25 nm.

Unexpectedly, the real part of the relative dielectric permittivity sharply increases from $10^3$ to $10^5$ with the temperature rise from -150°C to -50°C; then it quickly saturates to giant values ($\sim 5 \cdot 10^5$) and remains on the quasi-plateau in the broad temperature range (20 – 200)°C. The dielectric losses are very high ($tg\delta \sim 0.3 - 3$) and have a pronounced maxima in the region of the steep increase of the dielectric permittivity, which height increases slightly and position shifts strongly to the higher temperatures with the frequency increase. The temperature dependences of the electro-resistivity indicate the frequency-dependent transition in the electro-transport mechanisms between the lower and higher conductivity states accompanied by the maximum in the temperature dependence of the $tg\delta$.

The above-mentioned anomalous behavior of the dielectric response and losses of the SPS ceramic, which is observed in a broad frequency range (0.1 – 100) kHz, does not reveal any features of the ferroelectric-paraelectric phase transition near 125°C (as for the single-crystalline BTO) or at lower temperatures (as for the small BTO nanoparticles). The observed behavior is principally different from the typical behavior of the dielectric response of the HPS ceramics prepared from the same BTO nanoparticle powders, where we observed a pronounced maximum of the dielectric permittivity near 125ºC corresponding to the ferroelectric-paraelectric phase transition. At the same time, the pyroelectric thermal probing at the opposite surfaces of the SPS ceramic samples reveals the existence of the spatially inhomogeneous counter-polarized ferroelectric state. The state can be induced by thermal and/or conductive inhomogeneities. According to the SEM and Raman spectroscopy data, the inhomogeneities are graphite inclusions, which are non-uniformly distributed in the SPS ceramics.

Since the pyroelectric response proofs the presence of the spontaneously counter-polarized regions in the SPS ceramics, and since the dielectric measurements reveal the giant permittivity and high losses characteristic for the superparaelectric-like and semiconducting states, we assumed that the ceramics can be imagined as a strongly inhomogeneous media with electrically coupled semiconducting ferroelectric grains and much less conducting inter-grain. From the theoretical standpoint, the dielectric response of such inhomogeneous media can be described in the framework of effective media approach allowing for the Maxwell-Wagner effect. The giant superparaelectric-like dielectric response can appear in the semiconducting grains due to the well-known IBLC effect, as well as due to the step-like thermal activation of localized polarons at a definite temperature (which may be considered). However, since the experimental range of the dielectric measurements is below 1 MHz, it is not possible to specify the microscopic mechanism of the grain conductivity (e.g., IBLC [15-17] or polaron activation [13, 20]), which could be determined from the dispersion curves measured in the higher frequency range [18, 50, 51].

Obtained experimental results can be useful for the understanding of complex electrophysical properties inherent to the strongly inhomogeneous ferroelectric media.

**Acknowledgments.** Authors are very grateful to the Reviewers for Their great help in the manuscript correction and improvement. The samples preparation, characterization and results analysis are sponsored by the NATO Science for Peace and Security Programme under grant SPS G5980 "FRAPCOM" (S.I. and A.N.M.). A part of electrophysical measurements are sponsored by the Target Program of the National Academy of Sciences of Ukraine, Project No. 4.8/23-p (N.V.M., V.N.P., and V.V.V). Theoretical calculations performed by A.N.M. are funded by EOARD project 9IOE063b and related STCU partner project P751b. The work of O.S.P. and E.A.E. are funded by the National Research Foundation of Ukraine (projects "Silicon-compatible ferroelectric nanocomposites for electronics and sensors", grant application 2023.03/0127 and "Manyfold-degenerated metastable states of spontaneous polarization in nanoferroics: theory, experiment and perspectives for digital nanoelectronics", grant application 2023.03/0132).

**Author contributions**. O.S.P. conceived, performed (jointly with M.Y.Y., D.O.S. and O.V.B.) and analyzed results of the dielectric and electrophysical experiments (jointly with V.N.P.). S.E.I., B.P. and V.K. prepared the SPS ceramics and wrote Appendix A. A.S.N. performed Raman studies. V.I.S. performed the SEM measurements and wrote Appendix B. O.V.L. prepared the HPS ceramics. N.V.M. performed the pyroelectric measurements (jointly with M.Y.Y.) and analyzed results. A.N.M. and E.A.E. proposed analytical models of the dielectric response, performed numerical simulations and interpreted obtained results. The research idea, manuscript writing, and improvement belong to the corresponding authors, N.V.M., V.V.V. and A.N.M.

## Supplementary materials
## APPENDIX A. Samples preparation and characterization
### A1. Samples preparation

The BTO samples were obtained via the SPS using the furnace FCT HPD25 (FCT Systeme GmbH, Germany). Sintering was carried out in the graphite pressing tool with the die diameter of 20 mm using graphite foil to protect the die. Graphite provides effective Joule heating due to its good electrical conductivity. 5 g of BTO powder was preliminarily pressed under the load of 5 kN (16 MPa) in the same pressing tool in which sintering took place. Before starting sintering, a vacuum was created inside the chamber and the load of 16 kN (50 MPa) was applied. Heating was carried out at the rate of 400 °C/min, reaching 1100 °C and holding for 5 min. After the holding, the heating was immediately turned off and the load was gradually reduced to 5 kN (16 MPa) within 3 min. When the sample temperature reached 600 °C, the chamber was filled with nitrogen ($N_2$), and the load was finally removed. After that, the sample was left to cool down to room temperature naturally. The sintering mode of SPS of the BTO sample is shown in **Fig. A1**. After sintering, the sample with the diameter of 20 mm and height of 2.8 mm was obtained. The final density of the sintered sample was 5.99 g/cm$^3$ (99.5 %). The density was determined by the geometrical method, namely dividing the mass of the cut part of the sample by its volume.

The sintered sample has the inhomogeneous structure as shown in **Fig. A1**: the volume close to the edges is slightly darker than the inner volume. It may be due to the two following reasons. Firstly, carbon from the graphite tool can penetrate in the BTO sample and the highest concentration of carbon is close to the edges (see **Fig. A1**). But carbon can additionally penetrate deep into the entire sample as carbon-rich inclusions. Another reason is the inhomogeneous temperature and pressure distribution in the sample leading to the inhomogeneous distribution of the density. The temperature distribution in the sample depends on the temperature distribution in the graphite die. As a result, the highest temperature would be close to the edges. Moreover, temperature distribution may be especially inhomogeneous due to spatial gradients appeared at high heating rates.

To exclude the effect of "geometric" capacitance, which could appear in the highly porous mixtures of the insulating and conducting inclusions, we performed the following experiments. In the case of the effect of "geometric" capacitance the effective surface area of the capacitor could be much larger than the electrodes area, and the effective area could accumulate the space charge. We verified the geometric capacitance effect contribution by measuring the capacitance of the SPS BTO ceramic samples of different cutting angles, thickness, electrode area, frequency, and temperature, which does not reveal any scaling laws characteristic for the effect. Also, the microscopy observations do not confirm the high porosity of the studied samples. Hence, the geometric capacitance effect can be reliably excluded from the further consideration.

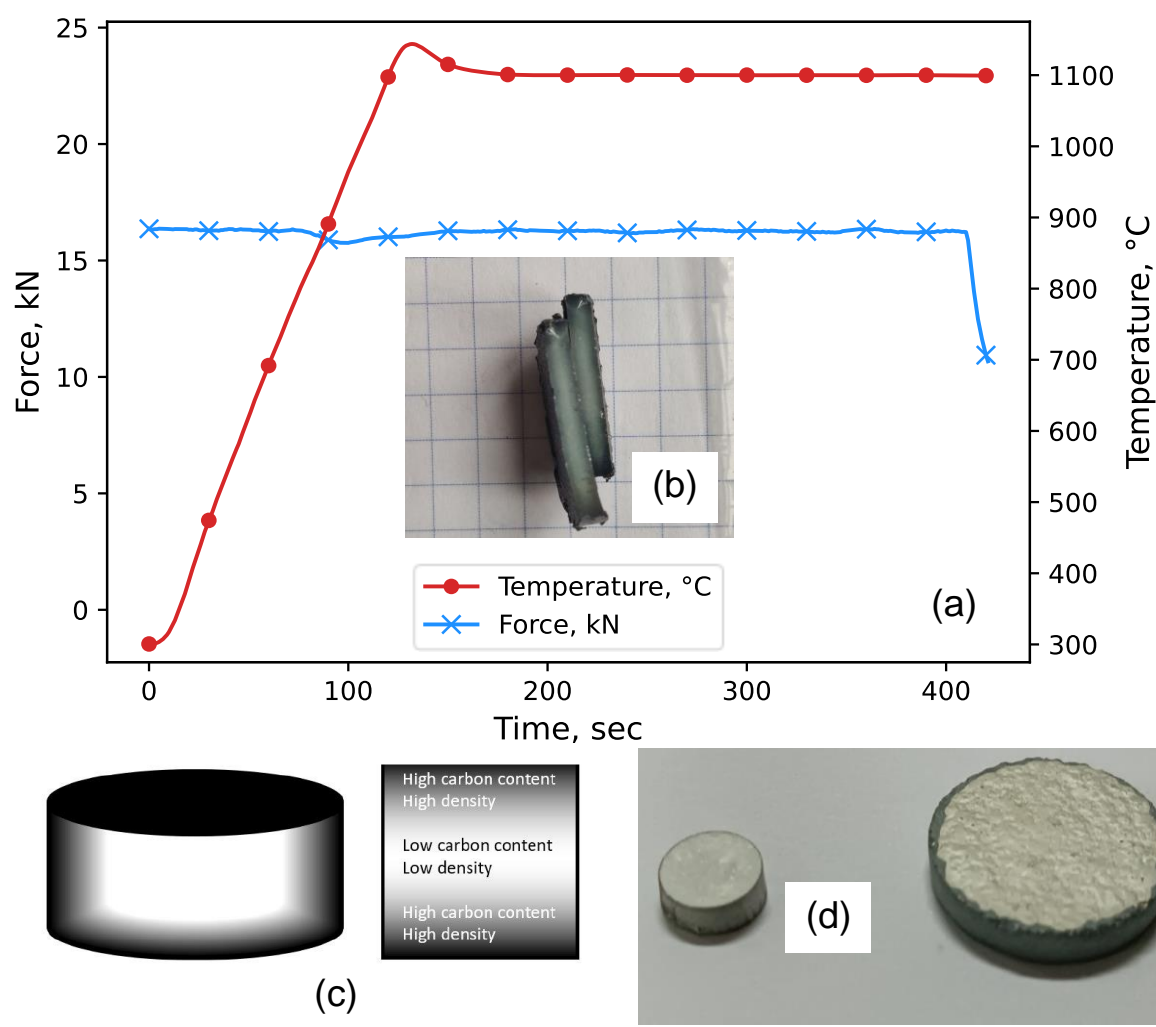

**FIGURE A1.** (a) Sintering mode of BaTiO$_3$ sample obtained via the SPS. (b) Fracture of the sintered sample. (c) Ununiform distribution of carbon in the sample. (d) Pellet annealed at 1250 °C by the HPS (left) and the pellet annealed by SPS at 1100 °C for 5 minutes, 400 °C/minute (right).

## A2. Samples characterization by the Scanning Electron Microscopy (SEM)

The BTO ceramics were studied in the JSM-35 scanning electron microscope. Two detectors of different types were used to obtain the image. They are the detector of slow secondary electrons and the detector of backscattered fast electrons. For the slow-electron detector, the brightness of the image is mainly determined by the topography of the sample (SEI contrast), while for the fast-electron detector, it is determined by the average atomic number Z of the area of the sample irradiated by electrons (Z contrast). However, for both detectors, the second component of the signal is also affected. By comparing these images, it was determined which feature of the sample relates to carbon. This is due to the large difference between the average atomic numbers of ceramics and carbon. For BTO, Z is 33.9, and for carbon it is 6. Therefore, graphite inclusions in the Z contrast mode look significantly darker. But other details of the relief, for example, holes or large pits, can also look dark. That is why the comparison of images was carried out. For example, the images below of one area were obtained using these two detectors. Graphite inclusions are marked by arrows on them. These images show that towards the middle of the sample, the number of carbon inclusions has decreased by about a factor of two.

At the same time, such studies are influenced by the operator's choice of the location of the image acquisition. Therefore, more comprehensive studies were additionally conducted. To do this, images in Z contrast of approximately half of the sample from the edge to its middle were obtained at a small magnification. After processing such an image, it was determined how the contrast Z signal changes from the edge to the middle of the sample. At the same time, a signal from twenty adjacent image lines was taken for averaging. The following figures show such an image and signal. When examining the sample, images of the areas were obtained with a step of 1 mm, starting from the very edge to its middle. These images obtained in Z contrast mode are presented in **Fig. A2.** It can be seen from the signal curve in **Fig. A3** that it increases from the edge of the sample to its middle, which corresponds to a decrease in the number of graphite inclusions in the ceramic.

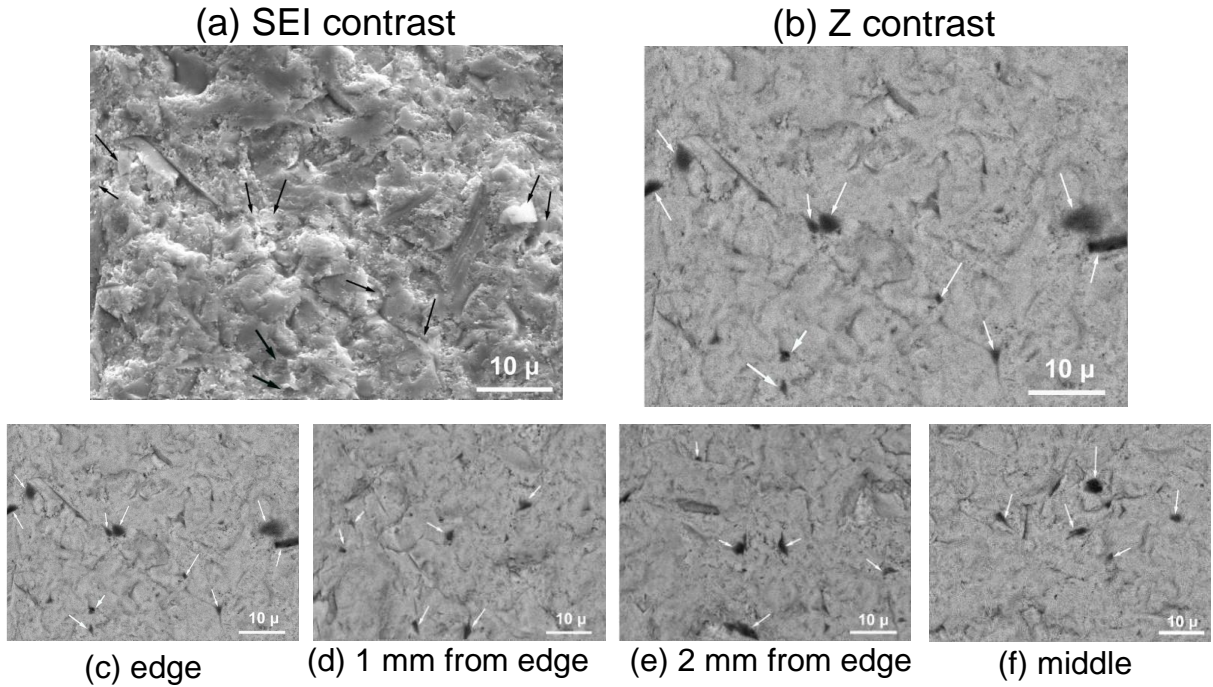

**FIGURE A2.** **(a)** SEI contrast and **(b)** Z-contrast images. Z- contrast images at the pellet edge **(c)**, 1 mm from the edge **(d)**, 2 mm from the edge **(c)** and in the middle of the pellet **(f)**. White arrows point on the black graphite inclusions.

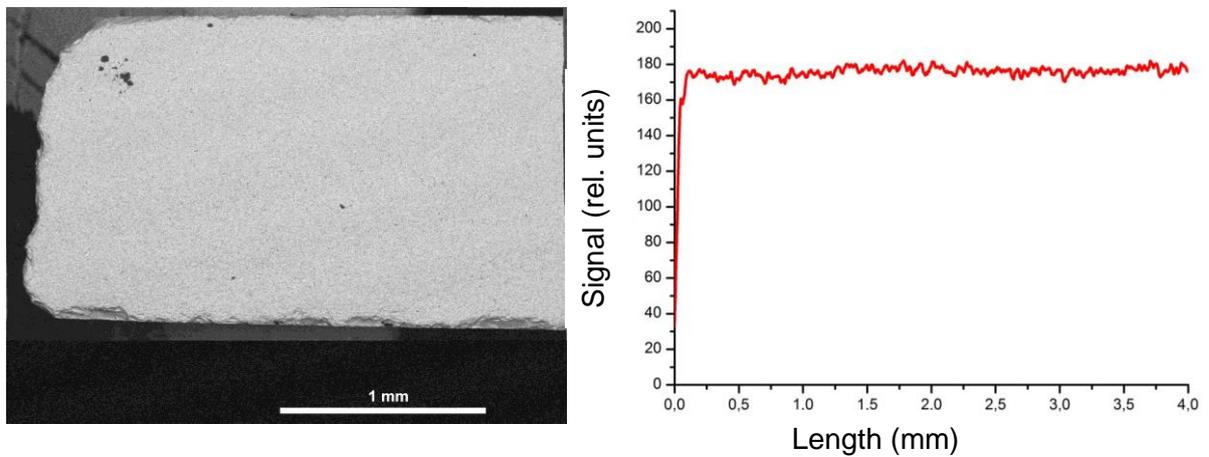

**FIGURE A3.** The SEM image of the sintered pellet **(a)** and the signal curve **(b)**.

## APPENDIX B. Maxwell-Wagner approach for ellipsoidal-like grains
### B1. Derivation of the equation for effective dielectric permittivity

According to the effective media model proposed by Liu et al. [57], the one can consider two (or more) layers representing all grain cores, their screening shells and grain boundaries, graphite inclusions and inter-grain space in the Maxwell-Wagner approach. One of the "effective" layers has a thickness $l$ and corresponds to the grain cores, and the other has a thickness $d$ and corresponds to all other regions (such as screening shells, grain boundaries and/or inter-grain space). The layers are characterized by the effective dielectric permittivity and conductivity: $\varepsilon_g$ and $\sigma_g$ for grain cores, $\varepsilon_{gb}$ and $\sigma_{gb}$ for grain boundaries and/or inter-grains. These effective parameters are temperature and/or frequency dependent. Assuming that the expression for capacitance of the series connection of the layers is valid, $\frac{l+d}{\varepsilon_{eff}} = \frac{l}{\varepsilon_g - i\frac{\sigma_g}{\omega}} + \frac{d}{\varepsilon_{gb} - i\frac{\sigma_{gb}}{\omega}}$, the effective permittivity of the SPS ceramics, $\varepsilon_{eff}$, is a complex function of the temperature $T$ and frequency $\omega$ of applied voltage, which can be presented in the form [57]:

$$\varepsilon_{eff}(T,\omega) = (l+d)\left(\frac{l}{\varepsilon_g - i\frac{\sigma_g}{\omega}} + \frac{d}{\varepsilon_{gb} - i\frac{\sigma_{gb}}{\omega}}\right)^{-1} \approx \varepsilon_\infty + \left(\varepsilon_s - \varepsilon_\infty - i\frac{\sigma}{\omega}\right)\frac{1}{1+i\omega\tau}. \quad \text{(B.1a)}$$

Here the following designations are introduced:

$$\varepsilon_\infty = \frac{l+d}{l/\varepsilon_g + d/\varepsilon_{gb}}, \quad \varepsilon_s = \left(\frac{\varepsilon_g}{\sigma_g} + \frac{\varepsilon_{gb}}{\sigma_{gb}}\right)\frac{l+d}{l/\sigma_g + d/\sigma_{gb}}, \quad \tau = \frac{l\varepsilon_{gb} + d\varepsilon_g}{l\sigma_{gb} + d\sigma_g}, \quad \sigma = \frac{l+d}{l/\sigma_g + d/\sigma_{gb}}. \quad \text{(B.1b)}$$

The complex dielectric permittivity in the right-hand side of Eq. (B.1a) consists of three contributions. The first term, $\varepsilon_\infty$, is the high-frequency limit of the dielectric permittivity determined by the dielectric permittivity of effective layers connected in series, namely $\varepsilon_{eff}(T, \omega \to \infty) \to \varepsilon_\infty$. The value of $\varepsilon_s$ is determined by the permittivity-to-conductivity ratios, as well as by the conductivity of the layers connected in series. The parameter $\sigma$ describes the contribution of the layers' conductivity. The time $\tau$ rules the timescale of the Debye-type relaxation determined by the conductivity and permittivity of the effective layers. The low-frequency limit of the effective permittivity (B.1a) real part is $\text{Re}[\varepsilon_{eff}(T, \omega \to 0)] \to \varepsilon_s - \sigma\tau$.

In the case when the morphology of the grains' connection is very complex, and the simplest model of series connection is not applicable, the expression for the effective permittivity should be derived. The derivation is based on the exact solution of the problem of ellipsoid with complex dielectric permittivity $\varepsilon_g^* = \varepsilon_g - i\frac{\sigma_g}{\omega}$ placed in the matrix with complex permittivity $\varepsilon_{gb}^* = \varepsilon_{gb} - i\frac{\sigma_{gb}}{\omega}$. The external electric field $E_{ext}$ has a fixed value far from the ellipsoid. It is well-known (see e.g., Ref.[58]) that the field inside the ellipsoid is equal to

$$E_g = \frac{\varepsilon_{gb}^*}{n_g \varepsilon_g^* + (1-n_g)\varepsilon_{gb}^*} E_{ext}. \quad \text{(B.2a)}$$

Here we supposed that the external electric field is pointed along one of the ellipsoid principal axes, for which the depolarization factor is $n_g$. The excess dipole moment $p_g$ of the ellipsoidal grain with semiaxes $a$, $b$ and $c$ is equal to:

$$p_g = \varepsilon_0 \frac{4\pi}{3} abc \frac{\varepsilon_g^* - \varepsilon_{gb}^*}{n_g \varepsilon_g^* + (1-n_g)\varepsilon_{gb}^*} E_{ext}. \quad \text{(B.2b)}$$

Below we use the approach proposed by Wagner [59] to derive the effective dielectric permittivity for the system of uniformly oriented dielectric ellipsoids of the same shape, separated far enough to neglect their mutual dipole-dipole interaction. Below we suppose that all grains are oriented along the external electric field, as well as the total dipole moment $m\,p_g$ of all the ellipsoidal grains should be equivalent to the total dipole moment of the virtual volume, which has the same shape and is filled with the matter of the effective permittivity $\varepsilon^*_{eff}$. An observer, who is unaware of the presence of the elliptical grains and who makes measurements on the substance, would find this apparent dielectric permittivity.

Let us calculate $\varepsilon^*_{eff}$ using the following consideration. Imagine a "big" ellipsoid with semiaxes $A$, $B$ and $C$, which contain a number $m$ of small ellipsoidal grains of the same shape. The space outside the "big" ellipsoid should contain the substance with permittivity $\varepsilon^*_{gb}$. In this space, therefore, at a sufficiently large distance from the "big" ellipsoid, the electrostatic potential is determined by the two contributions, namely those of the homogeneous field and the contribution of all the $m$ dipoles with the moments given by (B.2a), so that the total dipole moment is $p_{eff} = m \cdot 4\pi abc P_g/3$. If we now imagine that the virtual "big" ellipsoid with semiaxes $A$, $B$ and $C$ is filled by the substance with the effective permittivity $\varepsilon^*_{eff}$, the corresponding electric potential is determined by the presence of the dipole moment

$$p_{eff} = \varepsilon_0 \frac{4\pi}{3} ABC \frac{\varepsilon^*_{eff}-\varepsilon^*_{gb}}{n_g\varepsilon^*_{eff}+(1-n_g)\varepsilon^*_{gb}} E_{ext}. \tag{B.2c}$$

Instead of the number $m$ of small grains contained in the big ellipsoid it is more physical to use the ratio of the total volume of the small ellipsoid to the volume of the basic substance attributable to them, $\mu = m \cdot abc/(ABC)$, where the parameter $\mu$ has the meaning of the relative volume fraction of grains. Comparing (B.2a) and (B.2b), we get the following equation for $\varepsilon^*_{eff}$:

$$\frac{\varepsilon^*_{eff}-\varepsilon^*_{gb}}{n_g\varepsilon^*_{eff}+(1-n_g)\varepsilon^*_{gb}} = \frac{\varepsilon^*_g-\varepsilon^*_{gb}}{n_g\varepsilon^*_g+(1-n_g)\varepsilon^*_{gb}} \mu . \tag{B.3}$$

The solution of Eq.(B.3) with respect to $\varepsilon^*_{eff}$ is

$$\varepsilon^*_{eff} = \varepsilon^*_{gb} \frac{\{\mu+n_g(1-\mu)\}\varepsilon^*_g+(1-\mu)(1-n_g)\varepsilon^*_{gb}}{n_g(1-\mu)\varepsilon^*_g+\{1-n_g+\mu n_g\}\varepsilon^*_{gb}} = \varepsilon^*_{gb}\left[1+\frac{\mu(\varepsilon^*_g-\varepsilon^*_{gb})}{n_g\varepsilon^*_g+(1-n_g)\varepsilon^*_{gb}-n_g\mu(\varepsilon^*_g-\varepsilon^*_{gb})}\right] \tag{B.4}$$

It is seen that this expression gives $\varepsilon^*_{eff} = \varepsilon^*_{gb}$ and $\varepsilon^*_{eff} = \varepsilon^*_g$ for $\mu = 0$ and $\mu = 1$ respectively, as it should be expected.

For $n_g = 1$ (i.e., for the system consisting of the layers, perpendicular to the external field) one could reduce Eq.(B.4) to the following:

$$\varepsilon^*_{eff} = \left(\frac{1-\mu}{\varepsilon^*_{gb}} + \frac{\mu}{\varepsilon^*_g}\right)^{-1}. \tag{B.5a}$$

It is exactly the result for Maxwell's layered dielectric model.

For $n_g = 1/3$ (i.e., for the system consisting of the spherical grain) one rewrites (B.4) as

$$\varepsilon^*_{eff} = \varepsilon^*_{gb}\left[1 + \frac{3\mu(\varepsilon^*_g - \varepsilon^*_{gb})}{\varepsilon^*_g + 2\varepsilon^*_{gb} - \mu(\varepsilon^*_g - \varepsilon^*_{gb})}\right], \quad (B.5b)$$

which is the expression derived by the Wagner for the first time.

For $n_g = 0$ (i.e., for the system of the columns, parallel to the external field) one could get the following from of Eq.(B.4)

$$\varepsilon^*_{eff} = (1-\mu)\varepsilon^*_{gb} + \mu\varepsilon^*_g, \quad (B.5c)$$

which is equivalent to the system with parallelled capacitors of the same thickness but filled with the matters with permittivity values $\varepsilon^*_{gb}$ and $\varepsilon^*_g$.

### B2. The comparison of effective medium approximations

There are many effective media approaches (shortly "EMA"), among which the most known are the Landau approximation of linear mixture [60], Maxwell-Garnett [61] and Bruggeman [62] approximations for spherical inclusions, and Lichtenecker-Rother approximation of logarithmic mixture [63]. Most of these approximations are applicable for quasi-spherical randomly distributed dielectric (or semiconducting) particles in the insulating environment. The applicability of these EMA models is critically sensitive to the cross-interaction effects of the polarized nanoparticles, and therefore most of them can be invalid for dense composites and ceramics, where the volume fraction of ferroelectric particles is more than (20 – 30)%.

The effective media approximation proposed by Carr et al. [64] and later widely used (see e.g., Petzelt et al. [65] and Rychetský et. al. [66]), also abbreviated as "EMA", gives the quadratic equation for the effective permittivity of the binary mixture:

$$(1-\mu)\frac{\varepsilon^*_{eff} - \varepsilon^*_b}{(1-n_a)\varepsilon^*_{eff} + n_a\varepsilon^*_b} + \mu\frac{\varepsilon^*_{eff} - \varepsilon^*_a}{(1-n_a)\varepsilon^*_{eff} + n_a\varepsilon^*_a} = 0. \quad (B.6a)$$

Here $\varepsilon^*_a$, $\varepsilon^*_b$ are relative complex permittivity of the components "a" and "b" respectively, $\mu$ and $1-\mu$ are relative volume fractions of the components "a" and "b" respectively, and $n_a$ is the depolarization field factor for the inclusion of the type "a". The solutions of Eq.(B.6a) has the form:

$$\varepsilon^*_{eff} = \frac{(\mu - n_a)\varepsilon^*_a + (1-\mu-n_a)\varepsilon^*_b}{2(1-n_a)} + \sqrt{\frac{n_a\varepsilon^*_a\varepsilon^*_b}{1-n_a} + \frac{\left((\mu - n_a)\varepsilon^*_a + (1-\mu-n_a)\varepsilon^*_b\right)^2}{4(1-n_a)^2}} \quad (B.6b)$$

The other solution of Eq.(B.6a) with sign minus before square root is unphysical and should be excluded.

Hudak et al. [67] introduced a "single particle approximation" for the case of the small concentration of the ellipsoids with permittivity $\varepsilon^*_a$ distributed inside the matrix with the permittivity $\varepsilon^*_b$. They derived the expression for $\varepsilon^*_{eff}$,

$$\varepsilon^*_{eff} = \varepsilon^*_b\left[1 + \frac{\mu(\varepsilon^*_a - \varepsilon^*_b)}{n_a\varepsilon^*_a + (1-n_a)\varepsilon^*_b}\right], \quad (B.7)$$

valid in the case when $\mu \ll 1$. It is seen Eq.(B.7) is actually the first two terms of the expansion of the effective permittivity (B.6) with respect to $\mu$. Note that Bergman [68] also gave the derivation of the similar equations but with a different interpretation (he introduced the parameter $d = 1/n_a$ and called it as the dimensionality of the system).

Using the effective media approach, in the section B1 we derived linear equation (B.3) for the effective permittivity $\varepsilon_{eff}$:

$$\frac{\varepsilon^*_{eff}-\varepsilon^*_{gb}}{n_g\varepsilon^*_{eff}+(1-n_g)\varepsilon^*_{gb}} = \frac{\varepsilon^*_g-\varepsilon^*_{gb}}{n_g\varepsilon^*_g+(1-n_g)\varepsilon^*_{gb}}\mu \ . \tag{B.8}$$

Here we supposed that the external electric field is pointed along one of the ellipsoid principal axes, for which the depolarization factor is $n_g$.

The solution of Eq.(B.6) with respect to $\varepsilon^*_{eff}$ gives Eq.(B.4):

$$\varepsilon^*_{eff} = \varepsilon^*_{gb}\left[1 + \frac{\mu(\varepsilon^*_g-\varepsilon^*_{gb})}{n_g\varepsilon^*_g+(1-n_g)\varepsilon^*_{gb}-n_g\mu(\varepsilon^*_g-\varepsilon^*_{gb})}\right]. \tag{B.9}$$

It is seen that Eq.(B.9) contains the term, $-n_g\mu(\varepsilon^*_g - \varepsilon^*_{gb})$, in the denominator in comparison with the Hudak et al. Eq.(B.7). The term makes Eq.(B.9) consistent with the solution of the EMA equation (B.6).